# Strongly Coupled Electronic, Magnetic, and Lattice Degrees of Freedom in LaCo$_5$ under Pressure

Ryan L. Stillwell,[1] Jason R. Jeffries,[1] Scott K. McCall,[1] and Jonathan R. I. Lee,[1] Samuel T. Weir,[1] and Yogesh K. Vohra[2]

1. Materials Science Division, Lawrence Livermore National Laboratory, Livermore, California 94550, USA
2. Department of Physics, University of Alabama at Birmingham, Birmingham, Alabama 35294, USA

Abstract:

We have performed the first high-pressure magnetotransport and x-ray diffraction measurements on ferromagnetic LaCo$_5$, confirming the theoretically predicted electronic topological transition driving the magneto-elastic collapse seen in the related compound YCo$_5$. Our x-ray diffraction results show an anisotropic lattice collapse of the c-axis near 10 GPa that is also commensurate with a change in the majority charge carriers evident from high-pressure Hall effect measurements. The coupling of the electronic, magnetic and lattice degrees of freedom is further substantiated by the evolution of the anomalous Hall effect, which couples to the magnetization of the ordered state of LaCo$_5$.

Coupled degrees of freedom are at the root of the emergent behaviors of functional materials. Through their interconnected responses, lattice, electronic, magnetic, and orbital degrees of freedom can manifest physical effects ranging from superconductivity to ferroelectricity to magnetocaloric properties [1-4]. These effects are promising for myriad technologies including energy storage, quantum computing, and magnetic refrigeration, and materials exhibiting these effects have been the subjects of extensive research and development. In piezoelectric materials, an electric field (mechanical strain) develops as a result of applied mechanical strain (electric field) [5], and this coupling of strain and electric charge has allowed piezoelectrics to find use as sonar sensors, transducers, solar cells, and microantennas [6]. Additionally, multiferroic materials are being studied for a wide range of applications spanning compact spintronics-based data storage—where magnetic states can be manipulated with electric fields—to magnetic refrigeration through the magnetocaloric effect—where heat is absorbed through a magnetic-field-driven transition across a magnetic phase boundary [7, 8]. Coupled degrees of freedom clearly present promising properties, but a detailed understanding of the subtle couplings between lattice, electronic, and magnetic degrees of freedom is paramount to the continued development of functional materials.

While advances in computational models are improving the capabilities for predictive materials response, expanding the promise and potential of functional materials, there is still a critical need for experiments to validate theoretical frameworks. These experiments are naturally challenging, requiring capabilities to simultaneously assess seemingly disparate degrees of freedom under various conditions. Because it is intimately and thermodynamically connected to the volume of a material, and thus its crystallographic lattice, pressure is an effective tuning parameter for the exploration of functional materials. High pressures not only vary the solid lattice, and any degrees of freedom coupled to it, but modern, high-pressure techniques offer the capacity to interrogate the structure as well as the electronic and magnetic states arising from structural perturbations.

Pressure has been theoretically shown to generically cause a magneto-elastic collapse in compounds within the ferromagnetic $RECo_5$ system (RE – rare earth element including Y and La) [9], making these systems excellent platforms for experimentally studying the interplay between crystal, magnetic, and electronic structure. These $RECo_5$ compounds crystallize in a relatively simple hexagonal structure, and the magnetic RE ions (*e.g., not* Y, La, or Ce) interact with the crystalline electric field to orient the RE moment along a preferential crystallographic direction. On-site spin exchange between the RE 4f- and 5d electrons polarizes the latter, and inter-site hybridization between the RE 5d and the Co 3d states encourages the anti-parallel alignment of the RE and Co spins. For the light RE elements, the opposing spin and orbital moments of the RE ion results in a ferromagnetic arrangement of the RE and Co magnetic moments [10, 11]. This general mechanism is responsible for the properties of some of the most robust permanent magnet materials available today. A recent study implies that the $Sm_2Co_{17}$ system (a closely related structure to that of the $RECo_5$ system) should not exhibit a magneto-elastic collapse until pressures near 70 GPa[12]. On the other hand, $RECo_5$ compounds composed of non-magnetic RE are still ferromagnetic, but the absence of localized f-electrons may provide a less robust magnetic state and thus a propensity for magneto-elastic collapse at lower pressure. Indeed, electronic structure calculations predict an isostructural, magneto-elastic collapse in $YCo_5$ and $LaCo_5$ at 21 and 23 GPa, respectively [13]. According to the theory, an electronic topological transition drives this pressure-induced collapse when a majority-spin, occupied Co *3d*-band is forced across the Fermi level, resulting in a predicted reduction in ordered moment as well as a Fermi surface reconstruction [13]. While the structural signature of this magneto-elastic collapse has been observed experimentally in $YCo_5$ [14], no electronic or magnetic signatures of the collapse have been observed in either $YCo_5$ or $LaCo_5$.

Direct probes of changes in the magnetic moment or electronic structure are notoriously challenging experiments at high pressures. While spin-resolved photoemission would be sensitive to the predicted changes in the electronic structure of $LaCo_5$, the photoelectron energies are far too low to escape any containment used to generate pressures of order 10 GPa. Similarly, bulk magnetization and neutron scattering experiments at these pressures can suffer from extremely large backgrounds relative to the magnetic response of the sample. X-ray magnetic circular dichroism (XMCD) measurements at the Co K-edge—which require specialized facilities and extensive experimental time—would be sensitive to changes in the ordered Co moment with pressure and also amenable to high-pressure experiments [12, 15, 16], but quantitative measurement of the Co moment would be hampered by the inapplicability of spin-orbit sum rules at the Co K-edge. [17, 18] In a unique application of standard electrical conductivity measurements enabled through the use of designer diamond anvils, this article describes a multi-faceted suite of high-pressure measurements on $LaCo_5$ with which we are able to couple not only to the lattice, but also to the electronic and magnetic degrees of freedom at pressures in excess of 10 GPa. While x-ray diffraction measurements confirm the predicted isostructural collapse near 10 GPa, an associated, pressure-dependent magnetotransport study reveals the first experimental signatures of the magneto-elastic collapse in the magnetic and electronic channels.

Polycrystalline samples of $LaCo_5$ were prepared from a stoichiometric mixture of La and Co via arc melting, after which the sample was annealed at 1090°C for 28 days. Angle-dispersive x-ray diffraction measurements under pressure were performed using beamline 16 BM-D (HPCAT) of the Advanced Photon Source at Argonne National Laboratory. A gas-membrane-driven diamond anvil cell (DAC) was used to generate pressures up to 40 GPa. The sample was powdered, and loaded into the DAC sample chamber along with a Cu powder pressure calibrant; neon was used as the pressure-transmitting medium. Electrical transport studies under pressure were performed on a small polycrystal of $LaCo_5$ using an eight-probe designer DAC [19, 20] with steatite as the pressure-transmitting medium and ruby as the pressure calibrant. [21, 22] Longitudinal and transverse resistance measurements were performed as a function of temperature and magnetic field using the AC Transport option in a Quantum Design Physical Property Measurement System.

$LaCo_5$ crystallizes in the hexagonal $CaCu_5$-type structure (space group *P6/mmm*) with one formula unit per unit cell and the following atomic positions: La (1a), Co1 (2c), and Co2 (3g) [23]. Our x-ray diffraction results indicate that the $CaCu_5$-type crystal structure of $LaCo_5$ persists up to the highest measured pressures, near 40 GPa. However, while there are no structural transformations under pressure, the lattice does display an anomaly in the c-axis compression near 10 GPa. The diffraction patterns at each pressure were indexed to obtain the lattice parameters of $LaCo_5$ under pressure at ambient temperature (Fig. 1). The anomalous behavior in the c-axis compression can be seen in Fig. 1b (also see Appendix Fig. 1S) and contrasted with the conventional, anomaly-free pressure dependence of the a-axis as seen in Fig. 1a. The contraction is thus highly anisotropic. The c-axis contraction is small, amounting to a contraction of about 0.3% at 12 GPa. Because this c-axis contraction is small, the unit cell volume of $LaCo_5$ (Fig. 1c) does not show an obvious anomaly, permitting the use of a Birch-Murnaghan equation of state[24] to describe the system under compression ($B_0$=104 GPa, B'=4.3). In contrast to the volume, the c/a ratio shows a marked change in pressure dependence (inset Fig. 1c), exhibiting a pronounced flattening between 10 and 12.5 GPa.

The pressure-dependent c-axis anomaly and the flattening of the c/a ratio in $LaCo_5$ are very similar to the observed behavior seen in the isostructural compound $YCo_5$ [13, 14]. In the case of $YCo_5$, the anomalous structural behavior has been attributed to the effects of an occupied, majority-spin Co 3d-band crossing the Fermi level. When this d-band transects the Fermi level, some of the majority-spin, antibonding states are depopulated, resulting in a loss of magnetic moment and an increase in bond energy that contracts the lattice. Using the same magneto-elastic coupling scheme for $LaCo_5$, Koudela, et al. predicted a highly anisotropic lattice contraction with a 1.2% c-axis contraction occurring (within their calculations) at 7.7 GPa in the absence of an "overbinding" correction [13].

Our ambient-temperature experimental work is in excellent agreement with the highly anisotropic contraction predicted by DFT, but the magnitude of the c-axis contraction (0.3% vs. 1.2% for experiment vs. theory, respectively) and the pressure at which the transition occurs (10 GPa vs. 7.7 GPa for experiment vs. theory, respectively) show slight quantitative disagreement with predictions, likely arising from finite-temperature effects. Indeed, measurements of $YCo_5$ at ambient temperature show similar discrepancies with density functional theory calculations[13]. Altogether, the general agreement between previous theoretical predictions and our experiments on $LaCo_5$ lends credence to the magneto-elastic coupling scheme proposed by Koudela, et al., and Rosner, et al.[13, 14]

If the proposed magneto-elastic coupling scheme is a correct description of the physics of $LaCo_5$ at high pressure, then there should be commensurate changes in the electronic and magnetic degrees of freedom in $LaCo_5$ in the vicinity of 10 GPa as the Co 3d band crosses the Fermi level. To investigate this hypothesis we have performed a novel, high-pressure magnetotransport study enabled by an 8-probe designer diamond anvil (Fig. 2 inset) [20] that affords simultaneous probes of the *magnetic and electronic* channels of $LaCo_5$ through the magneto-elastic collapse. The designer diamond anvil permits measurement of both the longitudinal ($R_{xx}$) and transverse ($R_{xy}$) resistances versus pressure and magnetic field, and it is $R_{xy}$ in applied field that couples to both the magnetic and electronic degrees of freedom of $LaCo_5$.

In the absence of magnetism, $R_{xy}(H)$ is conventionally labeled as the Hall resistance after the Hall effect that describes its behavior. The Hall resistance is a function of applied field that is predicated on the electronic structure of a material, and thus any changes in electronic structure (*e.g.*, with applied pressure) will manifest as changes in the Hall resistance, making it a sensitive probe of the electronic degrees of freedom. In the case of ferromagnetic systems, $R_{xy}(H)$ will exhibit not only the conventional Hall resistance based on the electronic structure, but also the anomalous Hall effect (AHE). The AHE is directly coupled to the magnetic degrees of freedom. It is a result of the domains of a ferromagnet aligning with the applied magnetic field [25], and the magnitude of the AHE is proportional to the magnetization of the system, which also means that a non-zero transverse resistance can be realized in ferromagnets, even in zero applied magnetic field[12].

The total Hall effect in a ferromagnet is the sum of a "conventional" term and an "anomalous" term, such that the transverse resistance is expressed as

$$R_{xy} = R_H H_z + R_S M_z,$$

where $R_H$ is the conventional Hall coefficient, which is inversely proportional to the carrier density; $H_z$ is the applied magnetic field in the *z* direction; $R_s$ is the AHE prefactor, which is (generally) proportional to the resistance of the sample; and $M_z$, is the magnetization of the sample. There has been great debate on the subject of the proper form of the AHE prefactor.[26-32] All of the theories agree that $R_s$ is a function of $R_{xx}$ or the intrinsic band structure, yet the particular mechanisms and regimes where those mechanisms are dominant are still active areas of research [25]. The presence of these multiple scattering mechanisms, each of which may evolve differently with pressure, can make quantitative interpretation of the AHE signal challenging.

Figure 2a shows the anti-symmetrized (see SI for a detailed explanation of the anti-symmetrization procedure) $R_{xy}$ versus applied magnetic field up to 80 kOe, for pressures ranging from 2.1 to 15.6 GPa. The characteristic non-linearity of the AHE is evident at low field (<20 kOe), especially at the highest pressures. To isolate the anomalous part of the Hall effect we fit the slope of $R_{xy}$ over the applied magnetic field range 30-80 kOe, with the assumption that the slope above the saturation field (H~20 kOe) is solely due to the normal Hall effect. Subtracting the contribution from the normal Hall effect from $R_{xy}$ over the entire field yields the component due to the AHE:

$$R_{AHE} = R_{xy} - R_H H_z = R_s M_z,$$

where $R_{AHE}$ is the combined effect of the changes in the AHE prefactor and magnetization as a function of magnetic field. Figure 2b shows the extracted $R_{AHE}$ versus applied magnetic field.

To understand how the electronic structure of $LaCo_5$ changes as it goes through the isomorphic transition near 10 GPa, we plot both $R_H$ and the average saturation amplitude of $R_{AHE}$ (for 30<H<80 kOe), which we define as $R_A$, as functions of pressure in figure 3. Two interesting conclusions about the effects of pressure on $LaCo_5$ follow from the trends visible in Figure 2. First, when the normal Hall coefficient, $R_H$, is plotted as a function of increasing pressure (Fig. 3a), there is a sign change from negative to positive near 9.3 GPa, indicating a change in the dominant carrier type from electron-like to hole-like above this pressure. This change in $R_H$ is consistent with a Fermi surface reconstruction. Second, there is a concomitant increase in $R_A$ at the same pressure (Fig. 3b), demonstrating a change in the magnetic channel (*i.e.*, the product of the magnetization and AHE scattering prefactor) with pressure.

The simultaneous discontinuities observed in the electronic ($R_H$), magnetic ($R_A$), and lattice (c-axis lattice parameter) degrees of freedom with increasing pressure strongly suggest that the isostructural collapse of $LaCo_5$ is indeed induced by the predicted magneto-elastic coupling and its accompanying Fermi surface reconstruction. Within the framework of their calculations, Koudela, et al., predicted a reduction in the magnetic moment upon entering the collapsed structure[13]; however, this is not obviously borne out in our magnetotransport experiments, as the magnitude of $R_A$—which is proportional to the average saturation value of the product of $R_s$ and $M_z$—increases after crossing into the collapsed structure. Our attempts to isolate the magnetization, $M_z$, as a function of pressure by dividing $R_A$ by $R_{xx}(P)$ (see appendix Fig. 2S to for magnetotransport of $R_{xx}$ versus pressure) and $R_{xx}^2(P)$, did not produce a decreasing trend in $R_A$ (shown in appendix Fig. 3S). This could imply that the theory is incorrect in its calculation of the magnetic moment; but this explanation is unlikely, as the Co moment is already large and in the high-spin state at ambient conditions. A more likely scenario is that the AHE does not simply follow the longitudinal scattering dependence of the material but reflects the strongly correlated interplay between the electronic, magnetic and lattice degrees of freedom in $LaCo_5$.

In summary, we have confirmed the predicted high-pressure isomorphic transition in $LaCo_5$ using both x-ray diffraction and magnetotransport measurements. The highly anisotropic collapse of the c-axis was observed at 10 GPa, with a change of ~0.3% at 12 GPa, in very good agreement with previous zero-temperature DFT calculations that predicted an approximately 1.2% collapse near 8-10 GPa.[13] These theoretical calculations predicted that there would be magnetic and electronic transitions accompanying the observed anisotropic lattice collapse. Indeed, concomitant changes in the electronic structure and magnetic response were measured via high-pressure magnetotransport, which revealed a change in carrier type and a sharp increase in the amplitude of the anomalous Hall effect at 9.3 GPa. Magnetotransport measurements show promise as sensitive probes of magnetism under pressure. Though the measurements of the AHE under pressure could not unambiguously quantify the magnetic moment of $LaCo_5$, further progress into understanding and modeling the underlying mechanisms responsible for the AHE could enable routine, quantitative magnetic measurements at high pressure.

We graciously thank K. Visbeck for assistance with DAC preparation. This work was performed under LDRD (Tracking Codes 12-ERD-013, 14-ERD-041) and under the auspices of the US Department of Energy by Lawrence Livermore National Laboratory (LLNL) under Contract No. DE-AC52- 07NA27344. Portions of this work were performed at HPCAT (Sector 16), Advanced Photon Source (APS), Argonne National Laboratory. HPCAT operations are supported by DOE-NNSA under Award No. DE-NA0001974 and DOE-BES under Award No. DE-FG02-99ER45775, with partial instrumentation funding by NSF. Beamtime was provided by the Carnegie DOE-Alliance Center (CDAC). YKV acknowledges support from DOE-NNSA Grant No. DE-NA0002014.

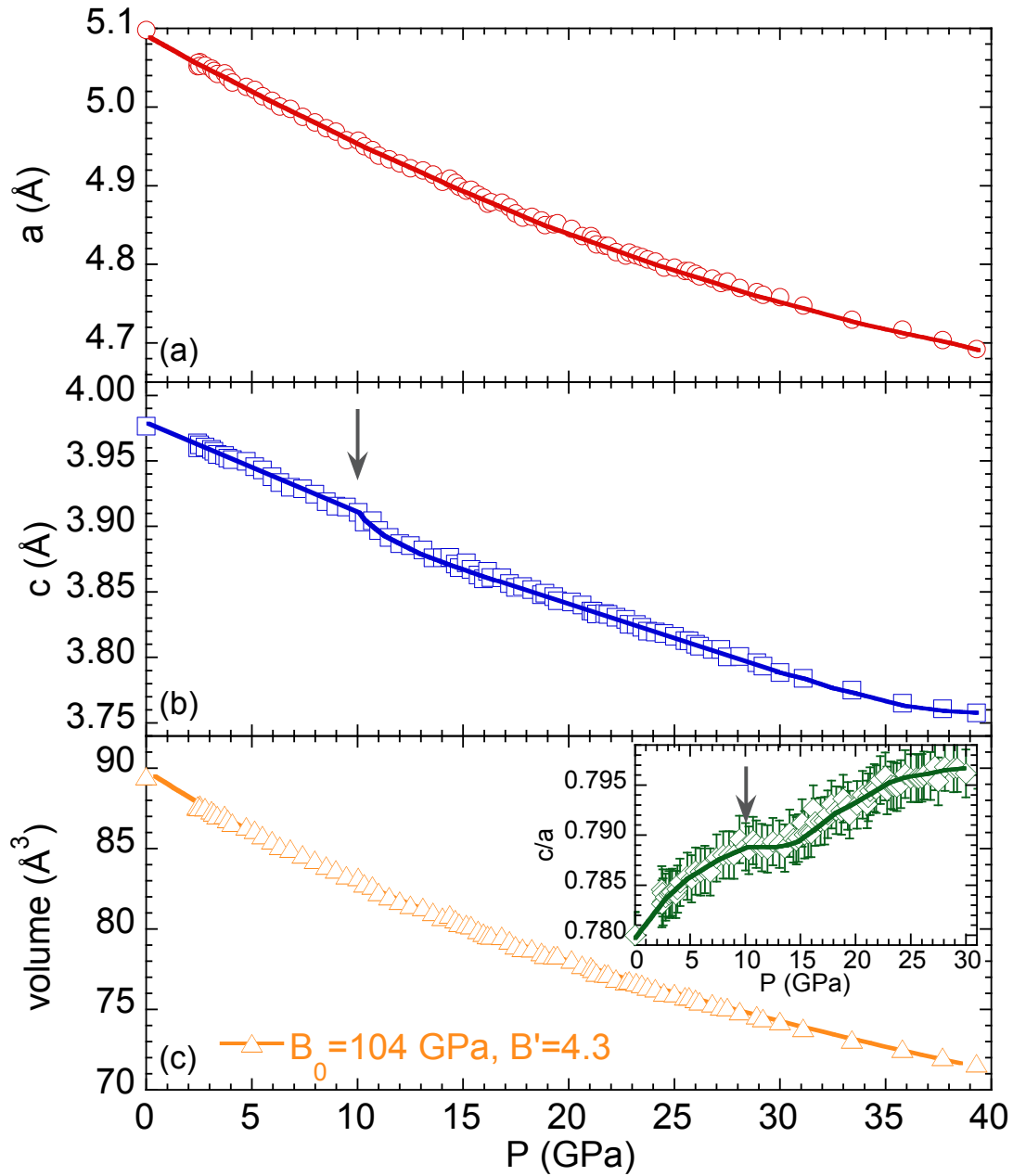

Figure 1 (color online). Hexagonal lattice parameters, a (a) and c (b), as well as the unit cell volume (c) for $LaCo_5$ under pressure at room temperature. While the a-axis compresses relatively smoothly, there is a marked deviation in the pressure dependence of the c-axis near the predicted isomorphic collapse near 10 GPa. Additionally, the pressure dependence of the c/a ratio, inset of (c), flattens near 10 GPa, before continuing to increase at higher pressures. Lines in (a), (b), and the inset of (c) are guides to the eye, while the line in (c) is a fit to the Birch-Murnaghan equation of state.

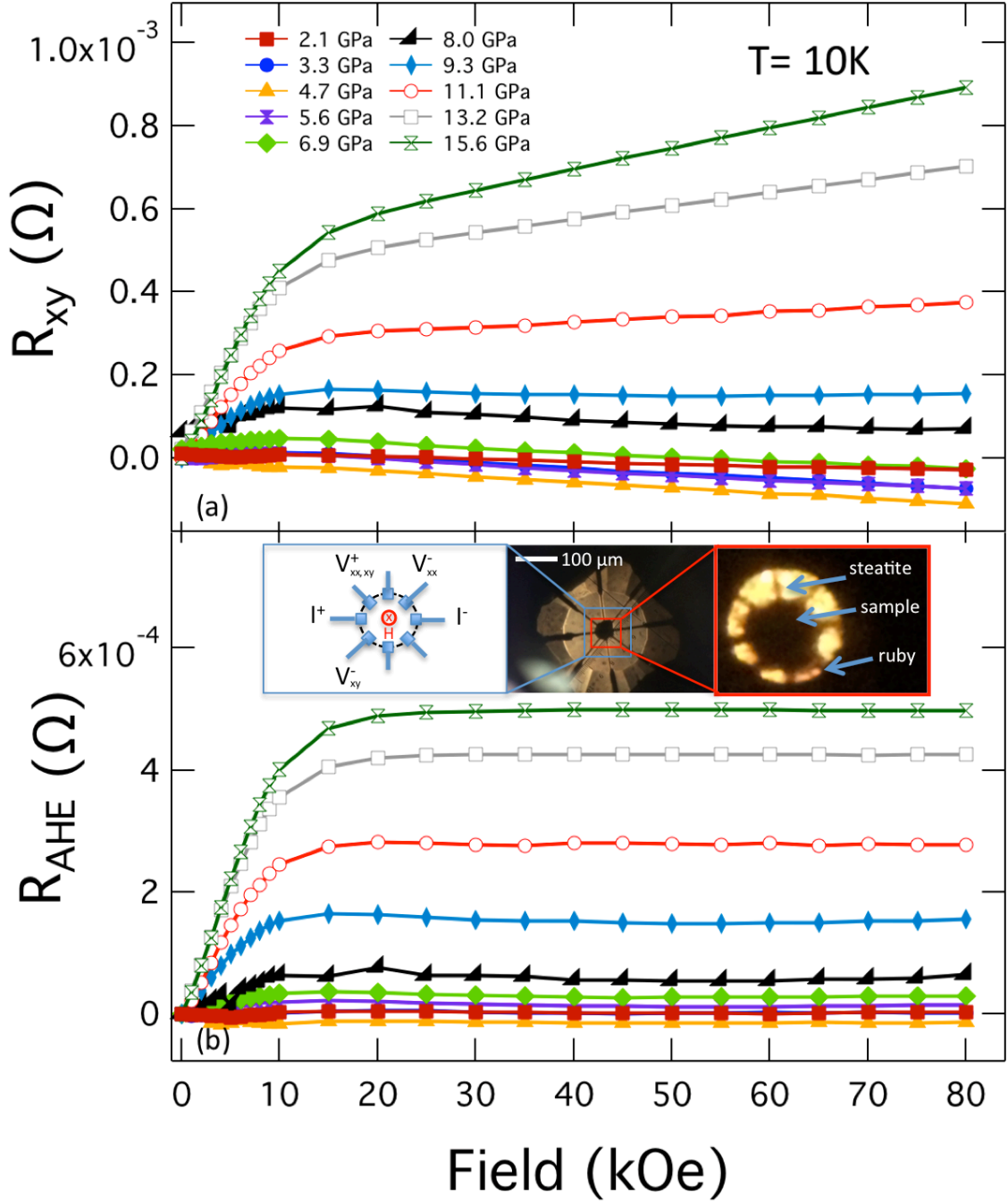

Figure 2 (color online). (a) Hall resistance of LaCo$_5$ as a function of increasing pressure, showing the increased amplitude of both the normal and anomalous Hall signals up to the highest pressure of 15.6 GPa. The normal Hall effect is defined as the slope from 30-80 kOe, after the saturation field where the ferromagnetic domains have aligned with the applied field (H~20 kOe). The Hall slope changes sign at 9.3 GPa (blue diamonds), also plotted as $R_H$ in Fig. 3(a). (b) The anomalous contribution is found by subtracting the normal Hall effect over the entire field range and is plotted as $R_{AHE}$ vs. magnetic field. This plot makes clear that the anomalous contribution also has an initial decrease, followed by an increase that grows rapidly from 9.3 GPa, concomitant with the isomorphic collapse seen in the X ray diffraction data shown in figure 1.

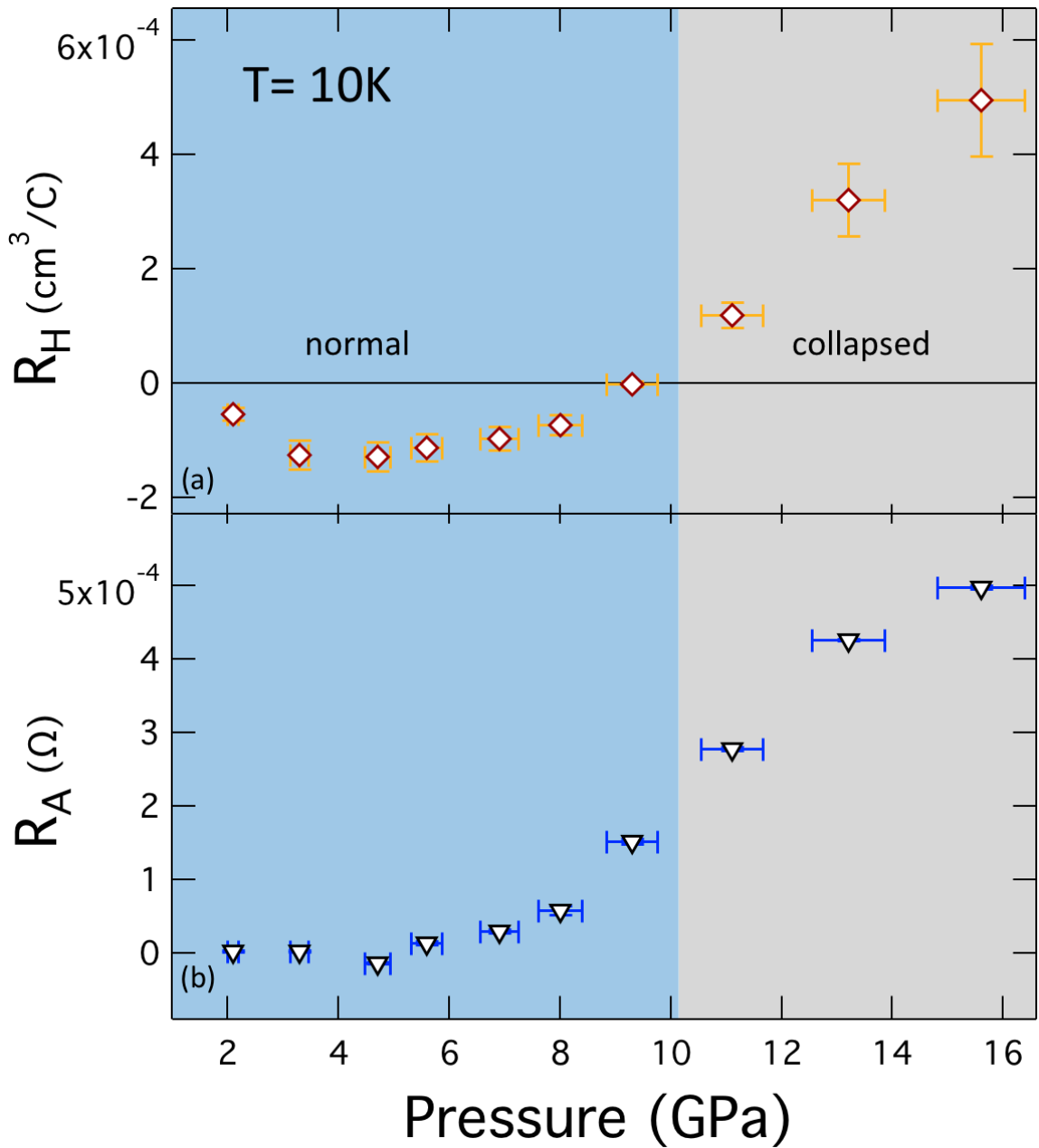

Figure 3 (color online). (a) Hall coefficient, $R_H$, of $LaCo_5$ plotted versus pressure up to 15.6 GPa (Error bars: $\Delta R_H$ is given as $R_H$ +/- 1 standard deviation of the fit to $R_{xy}$ between 30<H<80 kOe, $\Delta P$= P +/- 5%). The change in sign of $R_H$ at 9.3 GPa shows clearly that there is an electronic topological transition that accompanies the isomorphic structural transition from the "normal" to "collapsed" structure seen in our high-pressure x-ray diffraction study (Fig. 1). (b) $R_A$ plotted as a function of pressure up to 15.6 GPa showing a transition in the anomalous channel of the Hall signal reflecting changes in the magnetic degrees of freedom.